\documentclass[10pt,aps,pra,twocolumn,superscriptaddress]{revtex4}

\usepackage{dcolumn}
\usepackage[utf8]{inputenc}
\usepackage{lmodern}
\usepackage[T1]{fontenc}
\usepackage[english]{babel}
\usepackage[dvips]{graphicx}
\usepackage{enumerate,amsthm,amsmath,amssymb,color,graphicx,bbm,float}
\usepackage{bm}
\usepackage{mathtools}
\usepackage{natbib}
\usepackage{subfigure}
\usepackage{synttree}
\usepackage{overpic}
\usepackage{epstopdf}
\usepackage{tikz}
\usepackage{hyperref}
\usetikzlibrary{arrows}

\newcommand{\be}{\begin{equation}}
\newcommand{\ee}{\end{equation}}

\newcommand{\on}[1]{\operatorname{#1}}

\def\multiset#1#2{\ensuremath{\left(\kern-.3em\left(\genfrac{}{}{0pt}{}{#1}{#2}\right)\kern-.3em\right)}}

\begin{document}


\title{Robust Shot Noise Measurement for CVQKD}

\author{S\'ebastien Kunz-Jacques}\affiliation{SeQureNet, 23 avenue d'Italie, 75013 Paris, France}
\author{Paul Jouguet}\affiliation{SeQureNet, 23 avenue d'Italie, 75013 Paris, France}

\date{\today}

\begin{abstract}
We study a practical method to measure the shot noise in real time in Continuous Variable Quantum Key Distribution (CVQKD) systems. The amount of secret key that can be extracted from the raw statistics depends strongly on this quantity since it affects in particular the computation of the excess noise (i.e. noise in excess of the shot noise) added by an eavesdropper on the quantum channel. Some powerful quantum hacking attacks relying on faking the estimated value of the shot noise to hide an intercept and resend strategy \cite{LDG+:prl07} were proposed \cite{JKD:pra13}. Here, we provide experimental evidence that our method can defeat the saturation attack \cite{QKA:spie13} and the wavelength attack \cite{HKJ+:pra14}.
\end{abstract}

\maketitle

\section{Introduction}

QKD \cite{SBC:rmp09} is a quantum information technology primitive which allows two distant parties, Alice and Bob, to establish a secret key out of communications through an untrusted physical channel and a public authenticated channel. Its advantage over classical cryptographic primitives is that the produced keys are secure in the information-theoretic sense even against an eavesdropper limited only by the laws of quantum mechanics, i.e. in particular with unlimited computational resources and/or advanced knowledge of classical cryptanalysis techniques.

The first QKD realizations used unique photons to carry information. In their case, the impossibility to clone photons perfectly is the source of the security. Although the requirement to send unique photons was relaxed, their measurement still requires specialized detectors, usually cooled much below 0 degree Celsius. In contrast to this technology usually named Discrete Variable (DV) QKD, Continuous Variable QKD or CVQKD has been the subject of great interest in the past few years \cite{WPG:rmp12} because it can be implemented with only standard components optimized for the telecommunications industry and features nice characteristics such as a coherent detector which acts as a natural filter against classical Raman noise \cite{QZQL10} which is the main source of noise in WDM infrastructures where QKD signals coexist with intense signals at other wavelengths. Thanks to the possibility to encode more than one bit per pulse, it was shown recently that its secret key rate is expected to be competitive with DVQKD techniques assuming some reasonable clock rate improvements \cite{JEK:pra14}. The practical secure distance of CVQKD is 80 km \cite{JKL+:natphoton13} taking into account finite-size effects and several other imperfections \cite{JKDL:pra12}. It is currently limited by the technical noise of the system which is attributed to a potential eavesdropper. 

Quantum hacking against commercial DVQKD systems has been demonstrated in the past few years through powerful attacks, namely the time-shift attack \cite{ZFQ+08}, the phase-remapping attack \cite{XQL10}, and the remote control of single-photon detectors using tailored bright illumination \cite{LWW10}. Other attacks proposed against discrete-variable QKD systems include Trojan horse \cite{GFK+:pra06}, device calibration \cite{JWL11}, and wavelength dependent beamsplitter \cite{LWH+:pra11} attacks. The latter has been adapted to CVQKD \cite{HKJ+:pra14} while saturation attacks \cite{QKA:spie13} are CVQKD-specific. Generally, hacking of practical devices is a central topic in QKD since this is where the high theoretical security expectations of QKD meet reality.

For any QKD system, some statistics about the transmitted data are used to evaluate the amount of information that may be in possession of an attacker. For DVQKD protocols, the transmitted data are usually binary strings, and the eavesdropper information estimate only depends on the bit error rate between Alice's and Bob's data, and is zero for a null bit error-rate. For coherent-state CVQKD, the picture is more complex: the received data is always noisy because of a quantum noise term, the shot noise. The estimate of the eavesdropper information depends on the \emph{excess noise} which is the difference between the observed noise and the shot noise, and also on the observed correlation between the emitter and the receiver. The excess noise is typically at least one order of magnitude smaller than the shot noise.

One of the main implementation difficulties of CVQKD is to measure the shot noise with the required precision \cite{JKL+:natphoton13} to be able to compute the excess noise on the signal. The historical method to estimate the shot noise is to calibrate once-and-for-all the slope of the local oscillator to shot noise linear relation on the homodyne detection, and then to measure in real time the power of the local oscillator. It was shown in \cite{JKD:pra13} that this relationship is prone to changes over time, especially  under the influence of an attacker. In the case of a polarization dependent scheme, as the one pictured in Fig. \ref{fig:optical_scheme}, the shot noise is usually measured at the entrance of Bob before polarizations are separated; in this setting, this shot noise estimation technique is prone to shot noise overestimation when a polarization drift causes the optical power on the detection to be smaller than expected.

Because of the shortcomings of this shot noise measurement technique, another technique was proposed in \cite{JKD:pra13}. It attempts to separate the shot noise and the signal noise contributions in the total measured noise on signal by adding an active attenuation device on Bob's signal path, and measuring the total noise for two values of the attenuation ratio, exploiting the experimental observation that the contribution of the signal noise is proportional to the signal itself, hence should be proportional to the attenuation factor imposed to the signal. 

Unfortunately, a theoretical attack scenario against this procedure was proposed in \cite{HKJ+:pra14}. It works by exploiting defects in typical beam splitters to introduce specially crafted signals in both the signal path and the local oscillator path on Bob's side, which causes the attenuation-to-noise relation to become non-affine. The computations to evaluate the contributions of the shot noise and the excess noise in the total noise then yield wrong results, and with correct attack parameters this produces an underestimated excess noise and an overestimated key rate. It was remarked in \cite{HKJ+:pra14} that using not two, but three attenuation ratios was sufficient to thwart this attack. 

In this paper we present a more general shot noise / excess noise measurement technique using an active attenuation device on Bob's side, and show how it thwarts several known attacks on CVQKD systems. This proposal is based on an earlier idea suggested
in a patent made by SeQureNet \cite{KJ:patent2013}, and in \cite{JKD:pra13}. More importantly, we show that experimental data extracted from a real CVQKD setup are compatible with this countermeasure, i.e. real data are not incorrectly flagged as being caused by an attacker. As we explain in more details in this paper, our strategy allows to detect attacks that modify the linear relation between the variance of the signal and the variance on the noise, which is observed in the laboratory when no attacker is present. Some attacks do not modify this relation if they are correctly implemented, like Trojan horse attacks \cite{GFK+:pra06} for example. However, some passive countermeasures like the use of optical fuses or optical isolators can be used to defeat such attacks.

In Section \ref{sec:opticalsetup}, we present our optical setup and detail how we obtain our measurement results. We devote Section \ref{sec:practicalresults} to practical excess noise measurements and study three cases: Section \ref{sec:practicalresults_noattacker} gives experimental results in the absence of an attacker while Section \ref{sec:practicalresults_wavelengthattack} and Section \ref{sec:practicalresults_saturationattack} respectively include simulations of the behavior of the countermeasure when faced with the wavelength attack \cite{HKJ+:pra14} and the saturation attack \cite{QKA:spie13}.

\section{Optical Setup, Measurements}
\label{sec:opticalsetup}


\begin{figure*}[!ht]
\includegraphics[width=16cm]{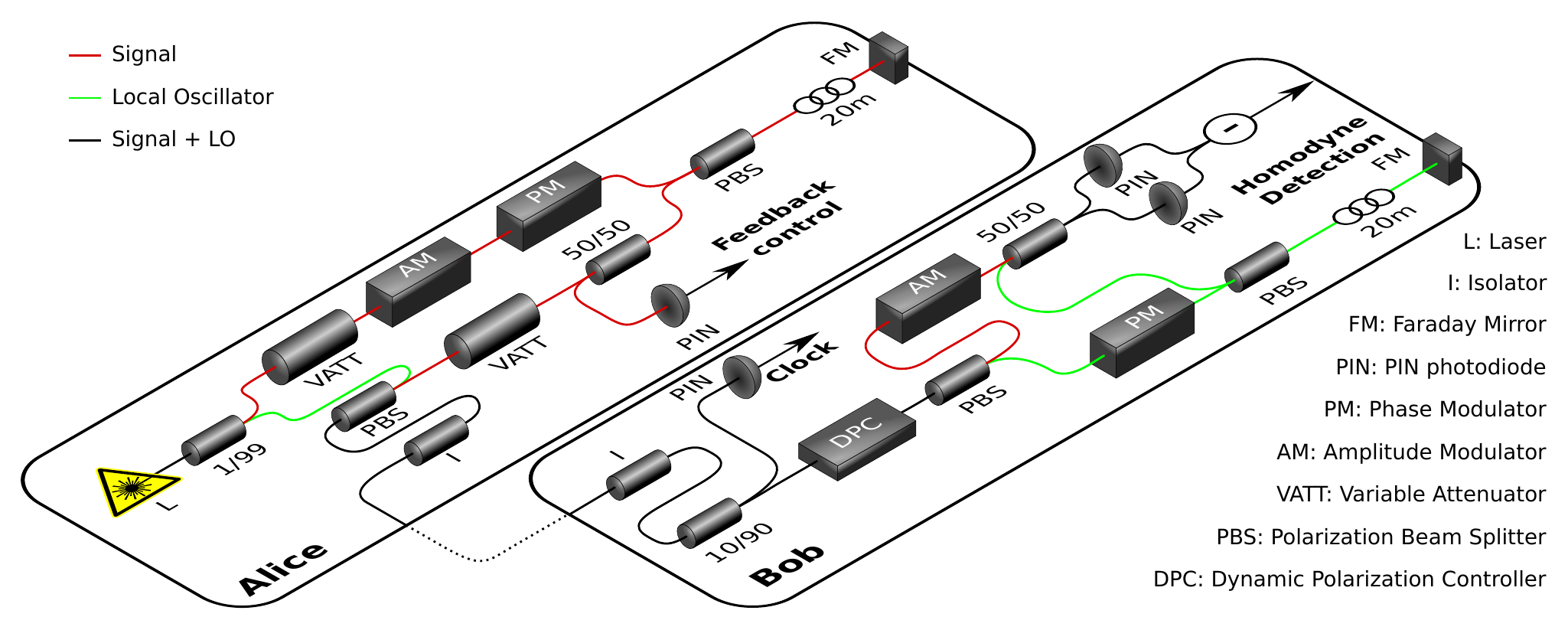}
\caption{\label{fig:optical_scheme} Optical scheme of a GG02 \cite{GG:prl02} implementation using time and polarization multiplexing, and including an amplitude modulator on the signal path of the receiver. Coherent light pulses with a duration of 100 ns and a repetition rate of 1 MHz are generated by a 1550 nm laser diode and split into a weak signal and a strong local oscillator (called the Local Oscillator or LO for short), which provides the required phase reference. Both the strong phase reference signal and the weak information-carrying signal are sent to Bob, thanks to a polarization and time multiplexing scheme: the signal pulse is 200 ns delayed with respect to the LO pulse using a 20 m delay line and a Faraday mirror while both pulses are multiplexed with orthogonal polarization using a polarizing beamsplitter (PBS). On Bob's side, the polarization drift that occurs on the quantum channel is actively compensated, then the LO and the signal are demultiplexed on Bob's side with another PBS combined with active polarization control. A second delay line on Bob's side allows for time superposition of signal and LO pulses. A phase modulator on the LO path enables to select the quadrature of the signal that is measured on the homodyne detection through the interference of the phase-modulated LO and the signal. An amplitude modulator is added on the signal path to simulate various signal attenuation ratios.}
\end{figure*}

Figure \ref{fig:optical_scheme} presents our implementation of the Gaussian coherent-state homodyne CVQKD protocol\cite{GG:prl02}. Alice modulates a Gaussian mixture of coherent states with a phase and an amplitude modulator. Coherent light pulses with a duration of 100 ns and a repetition rate of 1 MHz are generated by a 1550 nm laser diode and split into a weak signal and a strong local oscillator (called the Local Oscillator or LO for short), which provides the required phase reference. Both the strong phase reference signal and the weak information-carrying signal are sent to Bob, thanks to a polarization and time multiplexing scheme: the signal pulse is 200 ns delayed with respect to the LO pulse using a 20 m delay line and a Faraday mirror while both pulses are multiplexed with orthogonal polarization using a polarizing beamsplitter (PBS). On Bob's side, the polarization drift that occurs on the quantum channel is actively compensated, then the LO and the signal are demultiplexed on Bob's side with another PBS combined with active polarization control (the mean and variance of the measurements on the homodyne detection are used as feedback signals to compensate for slow polarization drifts using a Dynamic Polarization Controller). A second delay line on Bob's side allows for time superposition of signal and LO pulses. A phase modulator on the LO path enables to select the quadrature of the signal that is measured on the homodyne detection through the interference of the phase-modulated LO and the signal. An amplitude modulator is added on the signal path to simulate various signal attenuation ratios. For each attenuation ratio and for large enough groups of $N$ pulses, the measured signal $Y$ is projected as a $N$-dimensional vector on the signal $X$ emitted by Alice. Thus $Y$ is written $S + (Y-S)$
where $S$ is the signal proportional to $X$ and $Y-S$ is the noise orthogonal to $X$. In particular, when the signal path attenuation ratio is set to 0, the signal is 0 up to statistical noise and, assuming there is no manipulation of the LO, the shot noise is measured on the homodyne detection. Independently of the attenuation level, the noise $Y-S$ is equal to the sum of the shot noise and the excess noise. The current let through by the homodyne detection photodiodes is amplified by a transimpedance amplifier and the raw measurement results are therefore expressed in volts.

Assuming the homodyne detection has a linear signal response, the relationship of the excess noise variance to the signal variance, as well as the relationship between the signal variance and the attenuation ratio imposed on the signal path, should be linear. Indeed, any noise added to the signal will scale with the signal. A failure of one of these two tests gives a strong indication of the presence of an attacker as we shall see in the next sections.

Of course, to prevent any adaptive behavior of an attacker that would act differently depending on the attenuation ratio applied by Bob, it is required to choose randomly the attenuation ratio applied to individual pulses.  Furthermore the modulation of Alice must be independent of these attenuation choices. In our setup, these choices are made by Bob. Since secret key is produced only from the least attenuated groups, some information about the attenuation ratio selection is transmitted to Alice by Bob after the signal pulses are received. In the usual model where Alice's and Bob's laboratories are assumed to be out of reach of any attacker, this is sufficient to prevent any adaptive behavior of the attacker. Trojan horse attacks \cite{GFK+:pra06} are examples of attacks where Eve might acquire knowledge of Alice's and/or Bob's measurement and attenuation choices but these attacks, which break the usual assumption that Alice and Bob's device are out of reach of the attacker, are not considered in the present paper.

\section{Slope Measurements: practical Results}
\label{sec:practicalresults}

\subsection{Experimental Results Without an Attacker}
\label{sec:practicalresults_noattacker}

\begin{figure}[!ht]
\includegraphics[width=9cm]{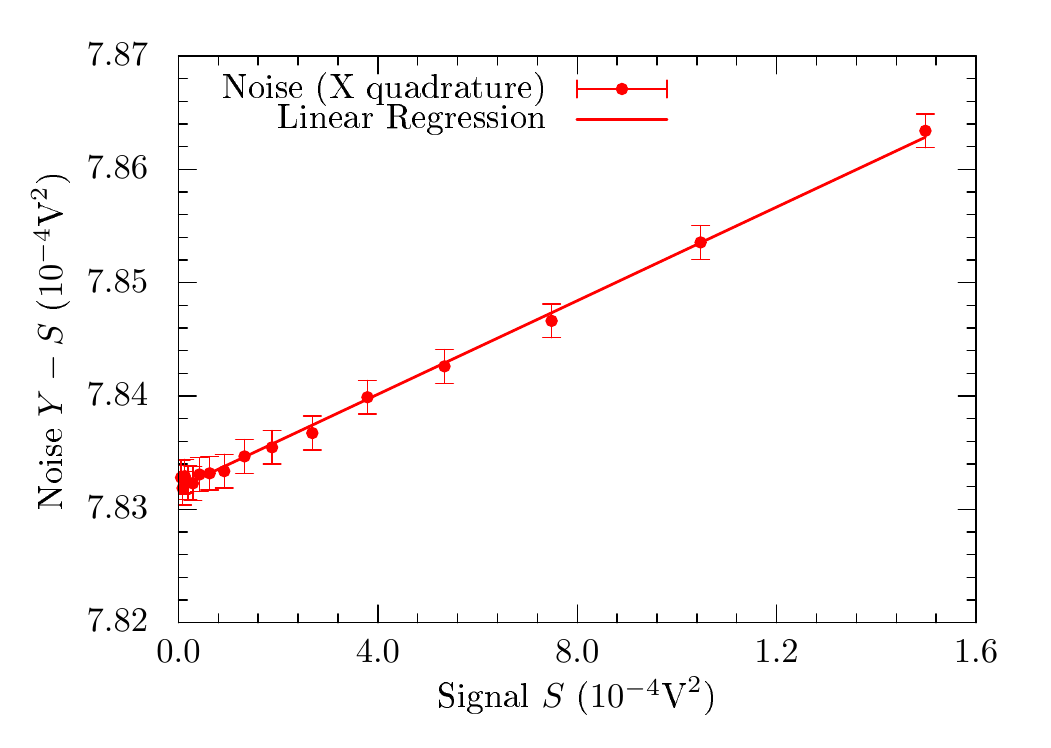}
\caption{\label{fig:signal_vs_noise_X} Noise variance as a function of signal variance in arbitrary units, X quadrature. The noise and signal variance are computed on approximately $N = 5 \times 10^8$ points in each group. The squared correlation coefficient $R^2$ is 0.9974 for quadrature X. The slope is equal to $2.07 \times 10^{-3}$. Error bars corresponding to 3 standard deviations of the statistical estimators are shown for the noise only, since for the signal variance they would be too thin to see.}
\end{figure}

\begin{figure}[!ht]
\includegraphics[width=9cm]{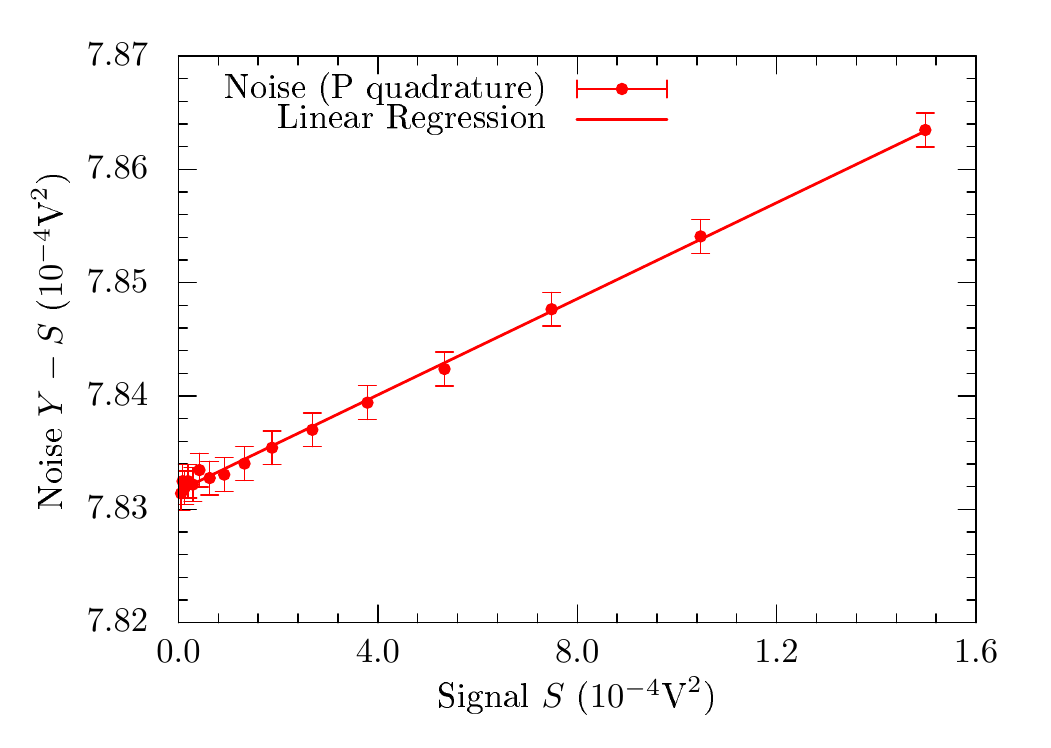}
\caption{\label{fig:signal_vs_noise_P} Noise variance as a function of signal variance in arbitrary units, P quadrature. The noise and signal variance are computed on approximately $N = 5 \times 10^8$ points in each group. The squared correlation coefficient $R^2$ is 0.9979 for quadrature P. The slope is equal to $2.12 \times 10^{-3}$. Errors bars are shown as in Fig. \ref{fig:signal_vs_noise_X}.}
\end{figure}

We first show the noise-to-signal and signal-to-attenuation relationships on a practical experimental setup. On Bob's side, 16 different attenuation settings are used with ratios $r_i$ satisfying $r_i/r_{i+1} \approx 0.7$. This ratio is chosen so that $r_0 / r_{15}$ is approximately equal to the amplitude modulator extinction ratio (here about 25 dB) in order to exploit its attenuation range. Data are gathered according to their attenuation setting and to the quadrature measurement performed ($X$ or $P$). The noise and signal variance in each group are computed on approximately $N = 5 \times 10^8$ points. The resulting signal-to-noise relationships are shown on Fig. \ref{fig:signal_vs_noise_X} and Fig. \ref{fig:signal_vs_noise_P}. Let $s_i$ be the signal variances and $n_i$ be the corresponding noise variances. The squared correlation coefficient $R^2=\frac{\on{Cov}^2(s,n)}{\on{Var}(s)\on{Var}(n)}$ which measures the quality of the affine correspondence between $s$ and $n$ is 0.9974 for quadrature X and 0.9979 for quadrature P. In shot noise units (i.e. divided by the constant term of the affine interpolation), the absolute values of errors w.r.t the linear model never exceed $1.3 . 10^{-4}$. This shows that a simple affine model for the noise-to-signal relationship can be observed in practice with good precision. The estimations of the shot noise derived by linear regression on both quadratures, despite being obtained from two different datasets, also agree with very good precision: on quadrature X, the estimate is 783.16 $\text{mV}^2$, while on P it is estimated to 783.19 $\text{mV}^2$, meaning that the relative error is about 3.5 $10^{-5}$. This error is consistent with our dataset size since it corresponds to approximately 0.55 standard deviations of the variance estimator of $N$ normalized Gaussian values. The slope of both curves, equal respectively to $2.07 \times 10^{-3}$ and $2.12 \times 10^{-3}$, show that the excess noises observed on Bob side for small SNRs (e.g. 0.1) are compatible with key production up to 100 km approximately.

\begin{figure}[!ht]
\includegraphics[width=9cm]{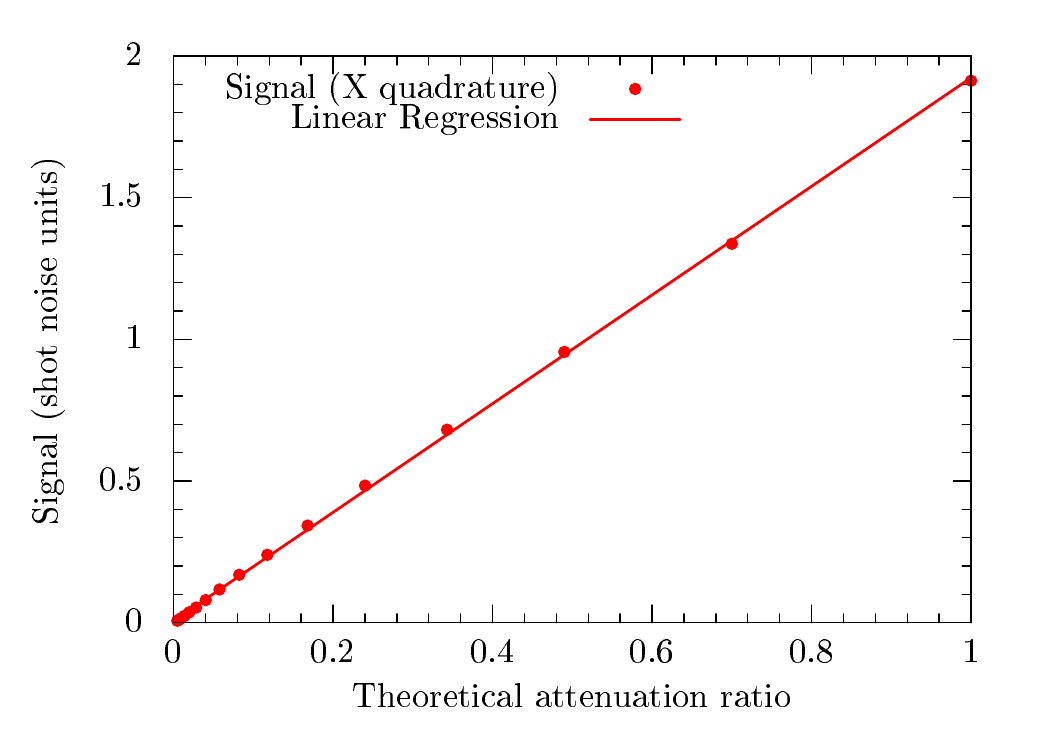}
\caption{\label{fig:signal_vs_att_X} Signal variance as a function of the theoretical attenuation ratio applied by Bob, X quadrature. The signal variance is computed on approximately $N = 5 \times 10^8$ points in each group. The correlation coefficient $R^2$ is 0.9996 for quadrature X.}
\end{figure}

Figure \ref{fig:signal_vs_att_X} shows the variation of the signal variance as a function of the attenuation applied by Bob, on the same dataset. Again we have a very good linearity ($R^2=0.9996$). Only the $X$ quadrature is represented, since  $X$ and $P$ curves are visually indistinguishable. Indeed, the small error term is not random but rather caused by a systematic error in the control of the attenuation ratio, and is very similar on $X$ and $P$ datasets.

One can ask why there is such a strong linear relationship between the excess noise variance and the attenuation applied by Bob when there is no attacker. Such a noise is for instance produced by an imperfect modulation, modeled by a noise term $Z$ added to the expected output of Alice $S$. This noise term is attenuated by Bob's active attenuation device exactly like the signal itself, and will contribute to the slope of the line in Fig. \ref{fig:signal_vs_noise_X} and Fig. \ref{fig:signal_vs_noise_P}. Conversely, a non-linearity of the homodyne detection would produce a noise term that is not proportional to the attenuation level. Our experiments therefore suggest that the main contribution to the technical noise of our system is caused by the imperfect modulation of Alice.

From these experimental results we derive the criteria that are applied to incoming data that govern whether data blocks are used to produce some secret key: When $16\times N$ data have been acquired at various attenuation levels, the least attenuated block is used for key production if the squared correlation factor $R^2$ for the linear fit of the noise-to-signal curve  is above 0.99 and the maximum error term in shot noise units (difference between the fitted line and the observed curve) does not exceed $2 . 10^{-4}$. Additionally the squared correlation factor of the attenuation-to-signal-variance relationship must be above 0.99. When some data is accepted, the excess noise and the shot noise used for the key rate estimation are computed with the help of the linear regression on the signal-to-noise curve. If secret key is produced only from the least attenuated group, 94\% of the pulses are used for parameter estimation. This can be overcome by using other blocks for key production, or by increasing the size of the least attenuated block. 

\subsection{Simulation with Wavelength Attack}
\label{sec:practicalresults_wavelengthattack}

\begin{figure}[!ht]
\includegraphics[width=8cm]{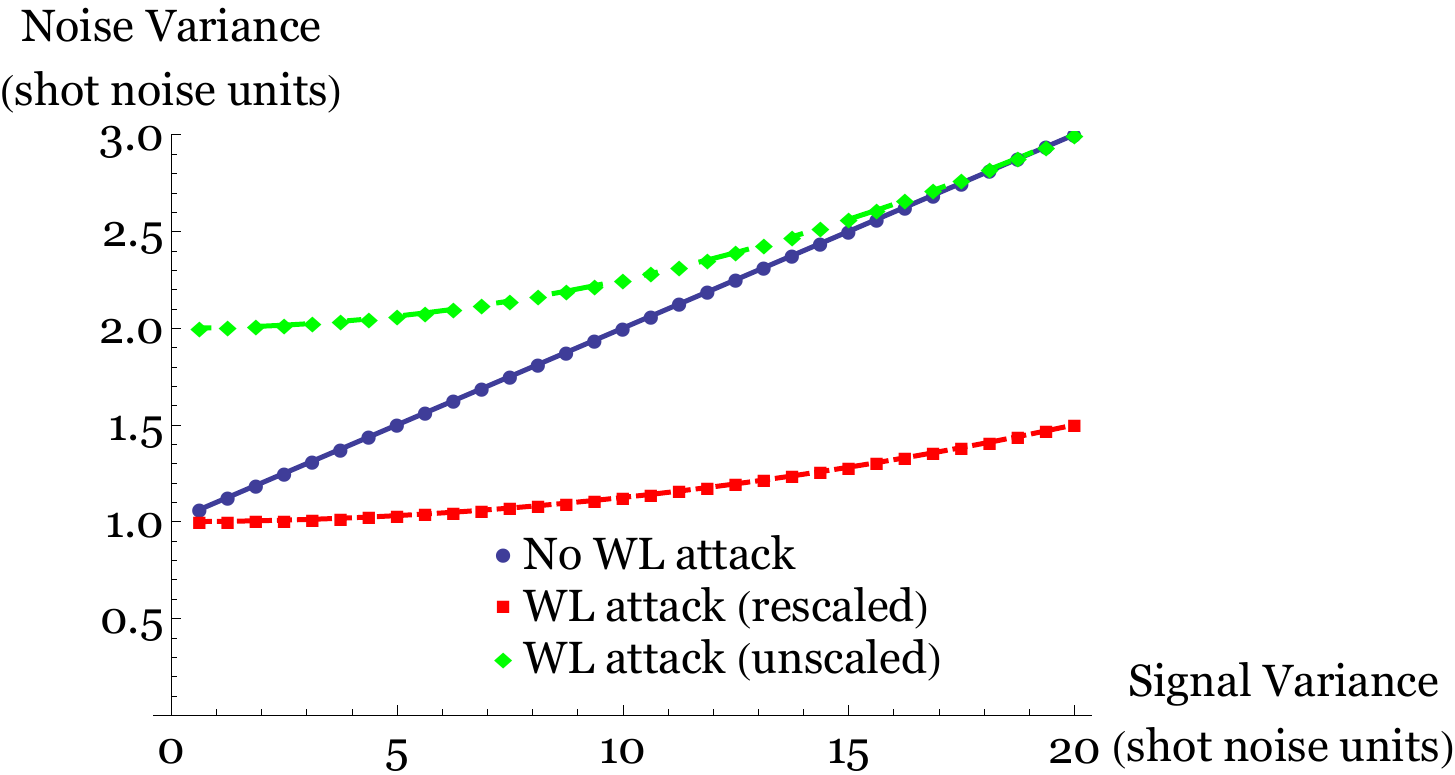}
\caption{\label{fig:signal_vs_att_WL_attack} Noise/Signal curve with and without wavelength attack. The non-normalized green curve shows the net effect of the attack on the detector: the added term is largest when the signal is 0, i.e. when the attenuation ratio is 0. After rescaling, the added term results in a lower estimate of the excess noise (red curve). Here $V_A=20$ and $T=1$.}
\end{figure}

In the wavelength attack \cite{HKJ+:pra14}, the optical setup and the imperfections of beam splitters are exploited to add to the measured values a deterministic signal whose amplitude depends on Bob's signal attenuation ratio as a degree-2 polynomial. The parameters of the attack can be chosen so that this polynomial has its minimum when the attenuation ratio is equal to 1. When the attenuation ratio is 0, a positive term is added to the shot noise variance and the sum of these two quantities is mistakenly used as the shot noise estimate, causing the excess noise to be improperly normalized. When combined with a full intercept-and-resend attack that is performed at the output of Alice, the noise added by the intercept-and-resend can be partially masked by the wavelength attack. An example of this scenario is summarized on Fig. \ref{fig:signal_vs_att_WL_attack}.

The signal-to-attenuation relationship is unaffected by this attack, hence its linearity cannot be used to detect the presence of the attacker. 

In the case of a full intercept and resend, the signal that must be added to cancel the excess noise induces a rather strong nonlinearity which results in a squared correlation factor of 0.94, and is therefore easily detected. In a more general scenario, the attack can be used to tamper slightly with the excess noise evaluation of the receiver. For a difference to linearity of $2 \,.\, 10^{-4}$ shot noise units as we observe normally in our setup, it is possible to use the wavelength attack to hide approximately $1\,.\, 10^{-3}$ units of excess noise on the receiver side for each unit of signal. This slope is below the technical noise of our system which is about $2\,.\, 10^{-3}$ units  of excess noise for each unit of signal as seen on Fig. \ref{fig:signal_vs_noise_X} and  Fig. \ref{fig:signal_vs_noise_P}. Therefore adopting the data acceptance criteria laid out at the end of section \ref{sec:practicalresults_noattacker}, the measured slope of the noise-to-signal relationship can be trusted up to $1\,.\, 10^{-3}$ units of noise per unit of signal variance. 

With this conservative parameter estimation procedure, the results of \cite{JKL+:natphoton13}, which showed a positive key rate at 80.5Km with a SNR of 0.075, remain valid, even taking into account the reduced efficiency $\eta \approx 0.3$ of Bob's detection device caused by the addition of an amplitude modulator on the signal path. This is in contrast to Fig. 5 of \cite{JKD:pra13}, which uses a higher noise figure as the one given by our new measurements. More precisely, a noise slope of $2 \, .\, 10^{-3} + 1 \,.\, 10^{-3}$, where the first term corresponds to the technical noise of our system and the second term is the margin of error possibly caused by the wavelength attack, leads to an estimated excess noise on Bob's side of $0.08 \times 3. 10^{-3} = 2.4 \,.\, 10^{-4}$ shot noise units. This corresponds at 80.5 Km to an excess noise on Alice's side of $2.4 \,.\, 10^{-4}\,.\,  10^{0.02 \,.\, 80.5}/0.322 = 3\%$. Alice's modulation variance needed to have a SNR of 0.075 on Bob side is equal to 9.5 units of shot noise. Reusing other parameters of \cite{JKD:pra13}, namely $\eta=0.322$, an electronic noise of 1\% of the shot noise and a reconciliation efficiency of 94.8\%, one gets a collective secret key rate of approximately $10^{-3}$ bits per symbol.

\subsection{Simulation With Saturation Attack}
\label{sec:practicalresults_saturationattack}

As the wavelength attack, the saturation attack  \cite{QKA:spie13} is aimed at hiding a full or partial intercept and resend by tampering with the excess noise measurement of the receiver. It assumes that the homodyne detection, when measuring a quadrature $x$ of some state, outputs not $x$ but $\operatorname{Min}(\operatorname{Max}(x, -\alpha), \alpha)$ for some $\alpha > 0$. Indeed, only a finite range of quadrature values can be adequately measured, either because of limitations of the analog electronics of the detection, or because of the range of the analog-to-digital conversion. To exploit this, the states received by Bob are shifted by an offset which causes the detection to be near the saturation point. When the offset is increased, both the signal and the noise variance tend to 0; indeed, for a sufficiently large value of the offset, the detection is stuck to its saturation level with overwhelming probability. If one uses an attenuation device on the signal path to shut off the signal and measure the shot noise, this  measure  is unaltered because the saturated signal is blocked when the shot noise is measured. As a result, a simple excess noise measurement where only the difference between the signal noise and the shot noise is computed may be defeated by this attack. However, simulations show that a more complete look at the signal and noise for several signal attenuations ratios easily reveals the saturation phenomenon.

All simulations are done with a value of $\alpha$ equal to 4 times the shot noise standard deviation, so that the shot noise variance measurement  is not significantly affected by the saturation limit of the detection. The offset imposed by the attacker is also equal to 4 times the shot noise standard deviation. A full intercept and resend attack is assumed, so that the excess noise is 2 units of shot noise for an unattenuated signal \cite{LDG+:prl07}.

\begin{figure}[!ht]
\includegraphics[width=8cm]{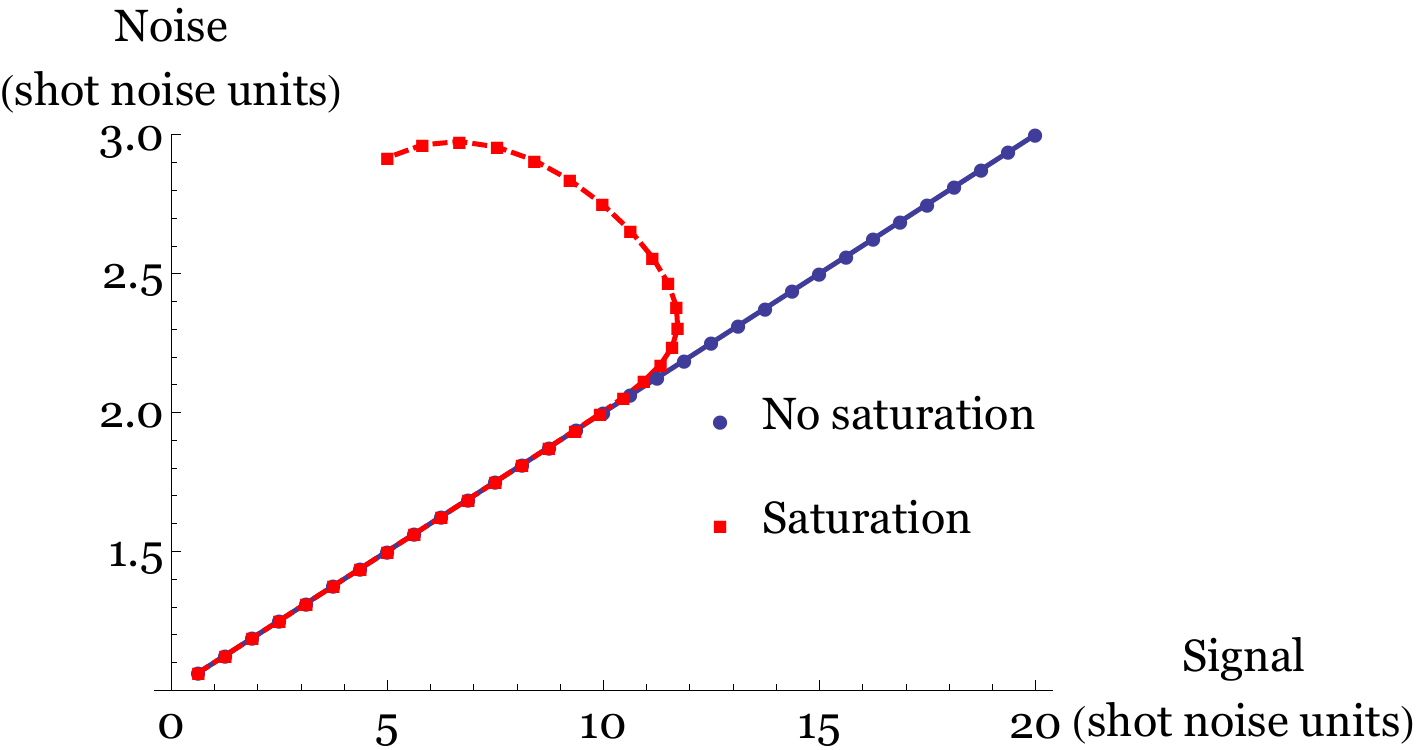}
\caption{\label{fig:sat_vb_20}Noise/Signal curve similar to experimental curves of Fig. \ref{fig:signal_vs_noise_X} and Fig. \ref{fig:signal_vs_noise_P}, with and without saturation, for a modulation variance on the receiver side $V_B=20$.}
\end{figure}

For a high variance on the receiver side, we obtain figure \ref{fig:sat_vb_20}. In this case the obtained curve is clearly not a line (the squared correlation coefficient is 0.386). Furthermore the apparent noise-vs-signal slope is higher than in the non-saturated case: in this regime, the attack does not improve the perceived excess noise slope.

\begin{figure}[!ht]
\includegraphics[width=8cm]{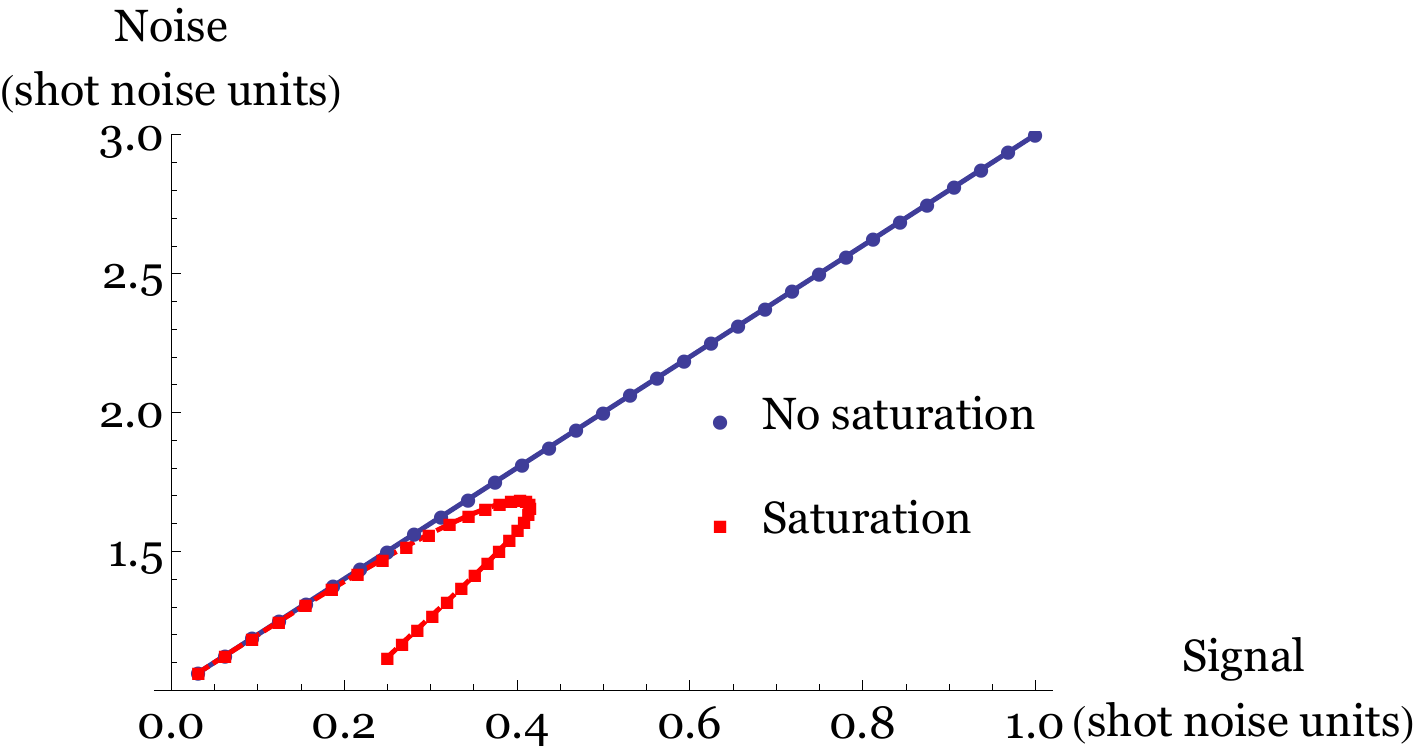}
\caption{\label{fig:sat_vb_1}Noise/Signal curve with and without saturation, for a line transmission of 1 and a modulation variance $V_B=1$.}
\end{figure}

On the contrary, for low variance on the receiver side such as the results shown in Fig. \ref{fig:sat_vb_1} (here a signal variance of 1 unit of shot noise), the noise variance falls before the signal variance falls too, and the resulting curve is below the normal unsaturated line, hence a direct comparison of the points with attenuation ratios of 0 and 1 would yield an excess noise slope lower than the real, unsaturated value. However the full curve observed is again very far from being a line (the squared correlation coefficient is now 0.665), which enables us to reject the acquired data without even considering the signal-to-attenuation response curve.

\begin{figure}[!ht]
\includegraphics[width=8cm]{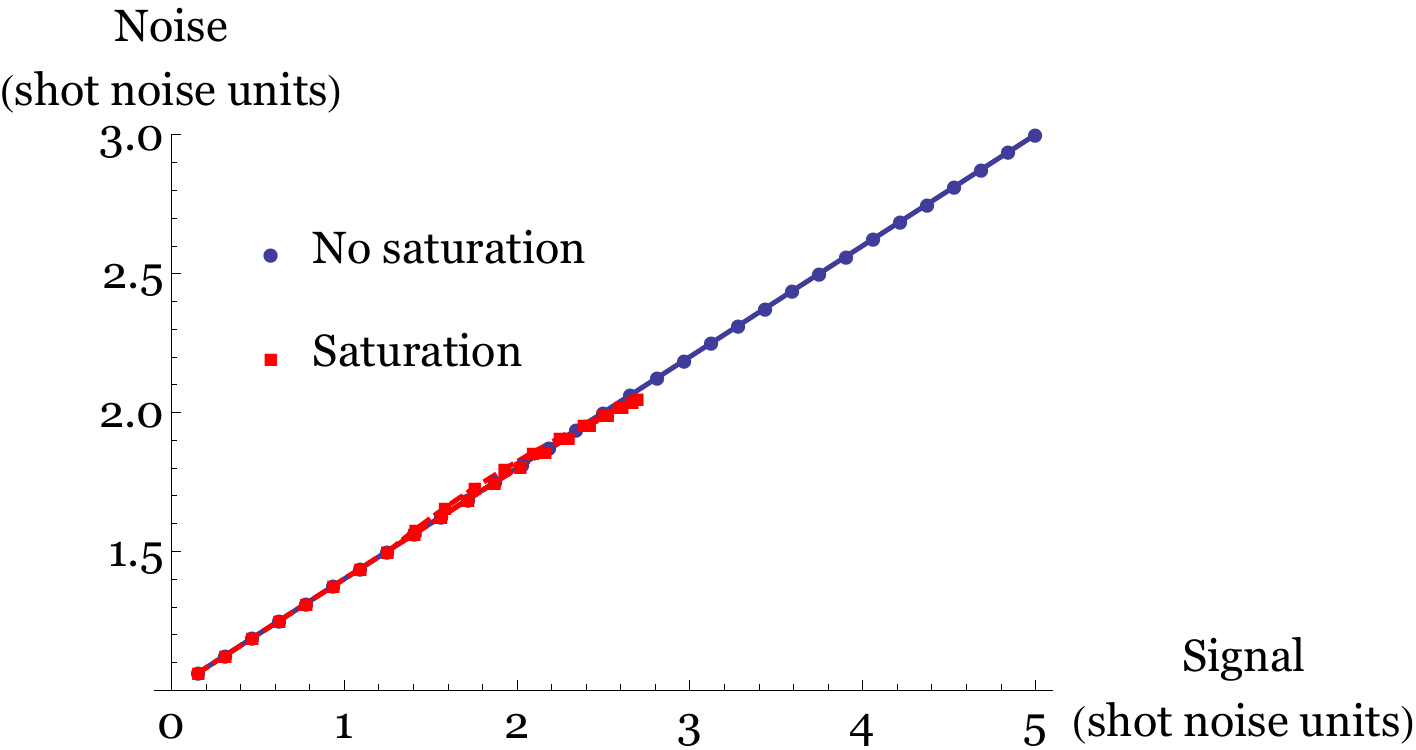}
\caption{\label{fig:sat_vb_5}Noise/Signal curve with and without saturation, for a line transmission of 1 and a modulation variance $V_B=5$.}
\end{figure}

\begin{figure}[!ht]
\includegraphics[width=8cm]{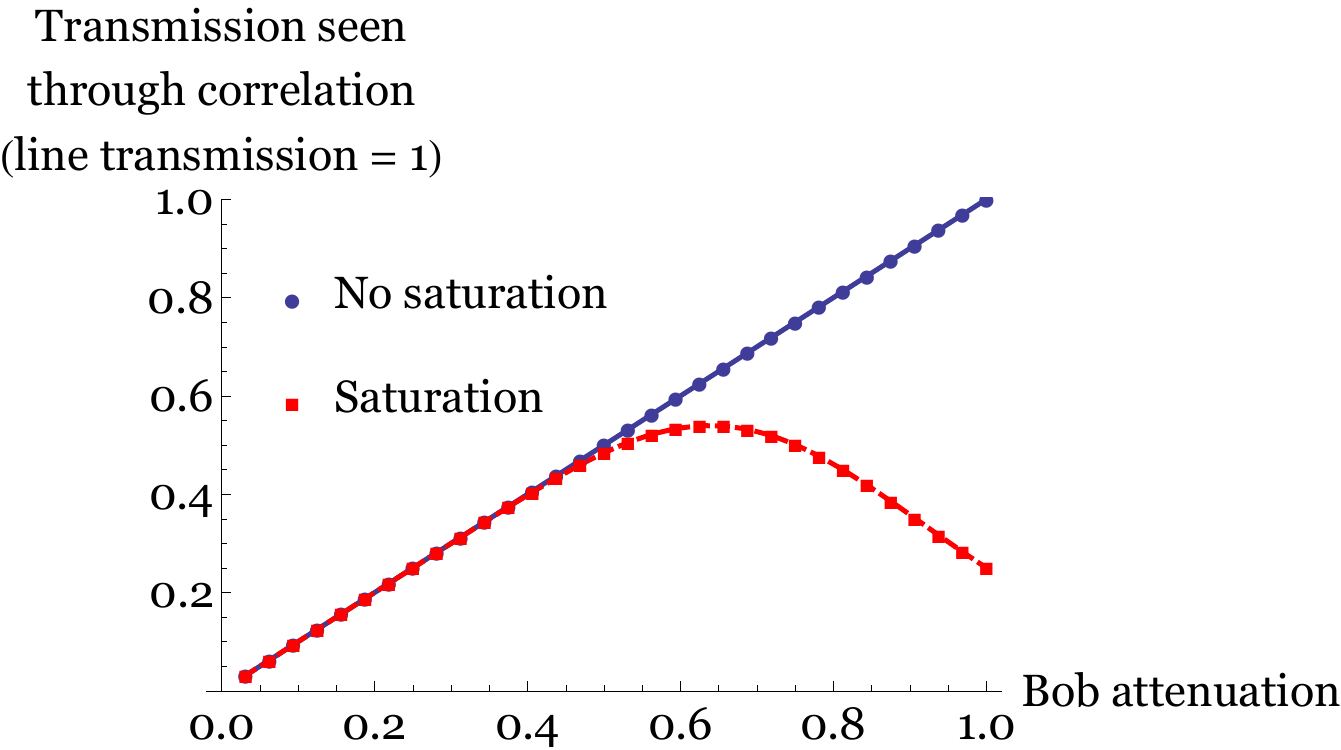}
\caption{\label{fig:sat_att_vb_5}Signal as a function of Bob's attenuation ratio with and without saturation, for a line transmission of 1 and a modulation variance $V_B=5$.}
\end{figure}

The limit case is represented by Fig. \ref{fig:sat_vb_5}. In that case, the noise and signal variance values, despite being severely distorted compared to the unsaturated case, still lie on the unsaturated line: the squared correlation coefficient is 0.998, with a slope very close to the unsaturated line (0.39 vs 0.40). Thus a direct look at this relationship may not be sufficient to reject the data. However, the signal-to-attenuation plot is clearly abnormal as shown on Fig. \ref{fig:sat_att_vb_5}: when the attenuation ratio is close enough to 1, the signal variance falls instead of increasing. 

Even without resorting to the signal-to-attenuation linearity check, it can be noted that the attack fails to lower significantly the excess noise slope, and therefore fails to hide the intercept-and-resend added noise in that case.

One area not explored in these graphs is the variation of the offset used by the attacker to cause the detection saturation. Its effect is to scale the saturated noise-to-signal curve: with a higher offset, the saturation effects appear for a smaller attenuation ratio. It does not change the general shape of the curve and in particular the squared correlation factor of the linear fit can only be improved by reducing the offset to reduce the overall saturation effect. We encourage attackers to experimentally test these theoretical dependencies and explore in details the behaviour of the homodyne detection as it has been done with single photon detectors \cite{LWW10}.

\section{Conclusion}

We proposed some specific criteria to decide whether data should be accepted for post-processing in an implementation  of Gaussian coherent CVQKD. These criteria depend on the normal behavior of an optical system implementing the protocol and define acceptable deviations from the ideal apparatus. It was shown that such a data filtering can defeat known attacks on coherent states CVQKD, except in the limit where the magnitude of the signature of the attack is comparable to the technical noise of the system. The parameter estimation can be modified in a conservative way to get around this limitation. Checking the linearity of the signal-to-attenuation response is an approximate way to test whether the homodyne detection output is proportional to a quadrature of the incoming electromagnetic field, which is an implicit hypothesis in all CVQKD security proofs. Because of the general nature of this test, we believe it applies to a wide range of potential attacks aimed at biasing the parameter estimation process and considerably improves the practical security of CVQKD.

\section{Acknowledgements}
This research was supported by the French National Research Agency, through the HIPERCOM (2011-CHRI-006) project, by the DIRECCTE Ile-de-France through the QVPN (FEDER-41402) project, and by the European Union through the Q-CERT (FP7-PEOPLE-2009-IAPP) project.

\bibliography{shot_noise_measurement}

\end{document}